\documentclass[a4paper,aps,pra,twocolumn,preprintnumbers,amsmath,amssymb]{revtex4-1}
\usepackage{amsmath}
\usepackage{verbatim}
\usepackage{graphicx,epstopdf}
\usepackage{hyperref}
\usepackage{soul}

\usepackage{color}

 \newcommand{\ket}[1]{\left|#1\right\rangle} 
 \newcommand{\bra}[1]{\left\langle#1\right|} 
  
\begin{document}
 
\title{Chiral spin currents in a trapped-ion quantum simulator using Floquet engineering}

\author{Tobias Gra{\ss}$^{1}$, Alessio Celi$^{2}$, Guido Pagano$^1$, Maciej Lewenstein$^{2,3}$}
\affiliation{$^1$Joint Quantum Institute, University of Maryland, College Park, MD 20742, U.S.A.}
\affiliation{$^2$ICFO-Institut de Ci\`encies Fot\`oniques, The Barcelona Institute of Science and Technology, 08860 Castelldefels, Spain}
\affiliation{$^3$ICREA, Psg. Llu\'is Companys 23, 08010 Barcelona, Spain}

\begin{abstract}
The most typical ingredient of topologically protected quantum states are magnetic fluxes. In a system of spins, complex-valued interaction parameters give rise to a flux, if their phases do not add up to zero along a closed loop. Here we apply periodic driving, a powerful tool for quantum engineering, to a trapped-ion quantum simulator in order to generate such spin-spin interactions. We consider a simple driving scheme, consisting of a repeated series of locally quenched fields, and demonstrate the feasibility of this approach
by studying the dynamics of a small system. An emblematic hallmark of the flux, accessible in experiments, is the appearance of chiral spin currents. Strikingly, we find that in parameter regimes where, in the absence of fluxes, phonon excitations dramatically reduce the fidelity of the spin model simulation, the spin dynamics remains widely unaffected by the phonons when fluxes are present. 
\end{abstract}

\maketitle
Magnetic fluxes can profoundly alter the behavior of a quantum system. They are responsible for the Aharonov-Bohm effect (\cite{bohm,Wilczek-book}), fractal energy spectra such as the Hofstadter butterfly (\cite{hofstadter,IQHE-book}), or intriguing many-body phases exhibiting the fractional quantum Hall effect (\cite{laughlin,Wen-book}). Most importantly, properties of a quantum system can be topologically protected in the presence of fluxes, as for example is the case for quantized Hall conductances. The underlying mechanism is due to breaking of time-reversal symmetry which fixes the chirality of the edge currents. Thereby, backscattering is suppressed, giving rise to robust values for the conductance. An exciting application of topological protection is fault-tolerant quantum computing using topologically protected states \cite{nayak}. In the present paper we will show that a magnetic flux can enhance the fidelity of a quantum simulation even in the few-body regime.

Quantum simulation of topological matter has been pursued very actively using cold atoms \cite{struck12,miyake13,aidelsburger13,jotzu14,Mancini15,stuhl15,meinert16}, photons in cavities \cite{hafezi13,schine16,roushan17,review_chiralqo_17}, or nitrogen-vacancies \cite{kong16}. Another experimental platform with long-standing history of quantum simulations involves trapped ions: Refs. \cite{mintert01,porras04} suggested the implementation of spin models, which have later been realized in the laboratory using linear Paul traps \cite{friedenauer08,kim09,kim10,richerme14,jurcevic14}, or two-dimensional Penning traps \cite{britton12,bohnet15}. Nowadays trapped ions are the leading platform for quantum simulation of spin models \cite{Zhang2017}. They offer high-efficiency detection and ion addressability at the single-site level \cite{naegerl99, Smith2016}, enabling full state tomography \cite{roos04} and  measurement of high order $N$-body correlators \cite{Jurcevic2017}.

A particular interesting feature of trapped ions is the long-range character of spin-spin interactions. This can be exploited to engineer magnetic fluxes even in a 1D architecture \cite{grass15}, for instance by having complex phases only on second-neighbor interactions. A standard route to artificial fluxes is Floquet engineering, that is the control of a Hamiltonian via periodic driving. Early proposals for modifying the tunneling strength in quantum wells using time-periodic fields date back to the 1980s and 1990s \cite{dunlap86,haenggi91,holthaus92,holthaus93}. Since then the progress in laser cooling and the emergence of cold atoms as highly controllable quantum systems have provided an experimental platform to apply these ideas. Controlling the tunneling of atoms in an optical lattice by periodic modulation was proposed in Refs. \cite{andre05,andre10} and realized in Refs. \cite{lignier07,struck11}. The technique then became a wide-spread tool for implementing artificial gauge potentials in atomic systems \cite{struck12,aidelsburger13,jotzu14,meinert16}. More generally, Floquet engineering has been recognized as a strategy for producing topological phases of matter such as topological insulators and Majorana fermions \cite{kitagawa10,jiang11,lindner11}. 

In this paper we study the feasibility of Floquet engineering in a system of trapped ions \cite{bermudez11,grass15,grass152}. In particular, we are interested in microscopic signatures of complex-valued spin-spin interactions. To this end, we consider a minimal system of three ions in a linear Paul trap. Similar to the scheme proposed in Ref. \cite{grass15}, a time-periodic series of local quenches is applied to the spins, that is, local potentials are repeatedly switched on and off. Such shaking protocol is able to approximately equalize all interaction strengthes, and to render interaction parameters complex. With spin-spin interactions having approximate XX symmetry (via the presence of a transverse field), 
the dynamics can be described as  a single particle (single hole) hopping along a triangle pierced by a magnetic flux. An immediate consequence of the flux, seen in the spin dynamics, is the chirality of the current.

While the description of the system in terms of effective spin model is rather simple and the effective Floquet Hamiltonian can be evaluated exactly, the real dynamics of three ions is actually much more complex. Indeed, apart from the spin dynamics, there are also vibrational degrees of freedom, which are only approximately decoupled as long as spin-phonon interactions are sufficiently fast. On the other hand, also the shaking of the spin model must be fast compared to the spin dynamics in order to get the desired Floquet Hamiltonian, but it should be slow compared to the spin-phonon coupling. To verify that this separation of time scales can be achieved, we have performed a simulation of the dynamics in the Dicke model, which includes also vibrational degrees of freedom. Strikingly, this simulation does not only show that Floquet engineering is feasible, but we also find that phonon effects, which reduce the fidelity of the quantum simulation, are suppressed through the presence of magnetic fluxes. This effect may be a signature of topological protection of chiral states in the full Hamiltonian governing phonons and spins. 

The starting point of our study is a Dicke-like Hamiltonian
\begin{align}
\label{dicke}
 H_0(t) = \sum_m \hbar \omega_m a_m^\dagger a_m + \sum_{i,m} \hbar \Omega_i \eta_{i,m} (a_m + a_m^\dagger) \sigma_i^x \sin(\omega t),
\end{align}
describing the collective motion of the ions in terms of phonons, created by $a_m^\dagger$ at frequencies $\omega_m$, coupled to two internal states of the ions. Note that we work in the (fastly) rotating frame of the atomic transition, and apply rotating-wave approximation.
Raman lasers with Rabi frequency $\Omega_i$ and beatnote frequency $\omega$ induce spin flips, described by a Pauli matrix $\sigma_i^x$ at an individual ion denoted by $i$, and excites or de-excites phonon modes. The strength of this coupling to the light further depends on the Lamb-Dicke parameters $\eta_{i,m}$. It is well-known \cite{mintert01,porras04} that, via a unitary transformation, the spin degrees of freedom can be decoupled from ``dressed'' phonon operators. By means of this decoupling, the time evolution of the ionic system for times longer than $2\pi/\omega$ is  captured by an effective Ising spin model \cite{kim09,wang12}, $H_J = \hbar \sum_{ij} J_{ij} \sigma_i^x \sigma_j^x$, with couplings $J_{ij} =  \Omega_i \Omega_j \sum_m \frac{ \eta_{i,m} \eta_{j,m}}{\delta_m}$, controlled by the detuning $\delta_m=\omega-\omega_m$.

For the shaking, we drive the system with a transverse magnetic field term $H_B(t) = \hbar \sum_i [B_0 + \mu_i(t)] \sigma_i^z \equiv H_{B0} + H_{B1}(t)$. The homogeneous and time-independent term $H_{B0}$ provides an approximate XX symmetry to the Ising model. The inhomogeneous time-dependent one, $H_{B1}(t)$ implements the shaking protocol. Note that on the level of the Dicke model, Eq. (\ref{dicke}), a transverse field leads to additional entanglement between spins and phonons \cite{wang12,Wall2017}. 

To derive the effective Hamiltonian, we note that $[H_{B0},H_{B1}]=0$. Thus, we may first switch to the interaction picture of $H_{B0}$ through the transformation $U(t)=\exp[-iH_{B0} t]$. Applying the rotating-wave approximation, we get an effective XX model, $U(t)^\dagger H_0(t) U(t)-i\hbar  U(t)^\dagger\frac{\rm d}{{\rm d}t} U(t)= H_{XX} + H_{B1}(t)$, with  $H_{\rm XX} = \hbar \sum_{ij} J_{ij} \,\sigma_i^+ \sigma_j^- +$ H.c., where $\sigma^{\pm}\equiv(\sigma^x\pm i \sigma^y)/2$. Then, we switch to the interaction picture of the time-dependent part $H_{B1}$, via $\tilde U(t)= \exp [ -i \sum_i \chi_i(t) \sigma_i^z ]$, where $\chi_i(t) = \frac{1}{\hbar} \int_0^t \ \mathrm{d}\tau \mu_i(\tau)$. The final Hamiltonian $H'_{XX}(t) = \tilde U(t)^\dagger H_{XX}(t) \tilde U(t)-i\hbar  \tilde U(t)^\dagger\frac{\rm d}{{\rm d}t} \tilde U(t)$ again has XX structure but with time-dependent couplings. By time-averaging over a period $T$, that is, in first order of the Floquet expansion, we get a time-independent XX Hamiltonian with effective couplings given by \footnote{Here, for simplicity we have assumed $B_0$ sufficiently larger than the Ising couplings $J_{ij}$ such that the net effect of driving is equivalent to the driving of a XX model by 
the time-dependent magnetic field $H_{B1}$. More generally, the condition to achieve an effective XX model with the complex coupling as in Eq. \eqref{jeff} is $\int_0^{\tilde T} \exp [2 i  (2 B_0 t + \mu_i(t) +\mu_j(t))] =0$, where
$\tilde T$ is the period of $ B_0 t + \mu_i(t)$.}
\begin{align}
\label{jeff}
 J_{ij}' = \frac{J_{ij}}{T} \int_0^T   e^{2i[\chi_i(t)-\chi_j(t)]}{\rm d}t.
\end{align}

\begin{figure}[t]
\centering
\includegraphics[width=0.48\textwidth, angle=0]{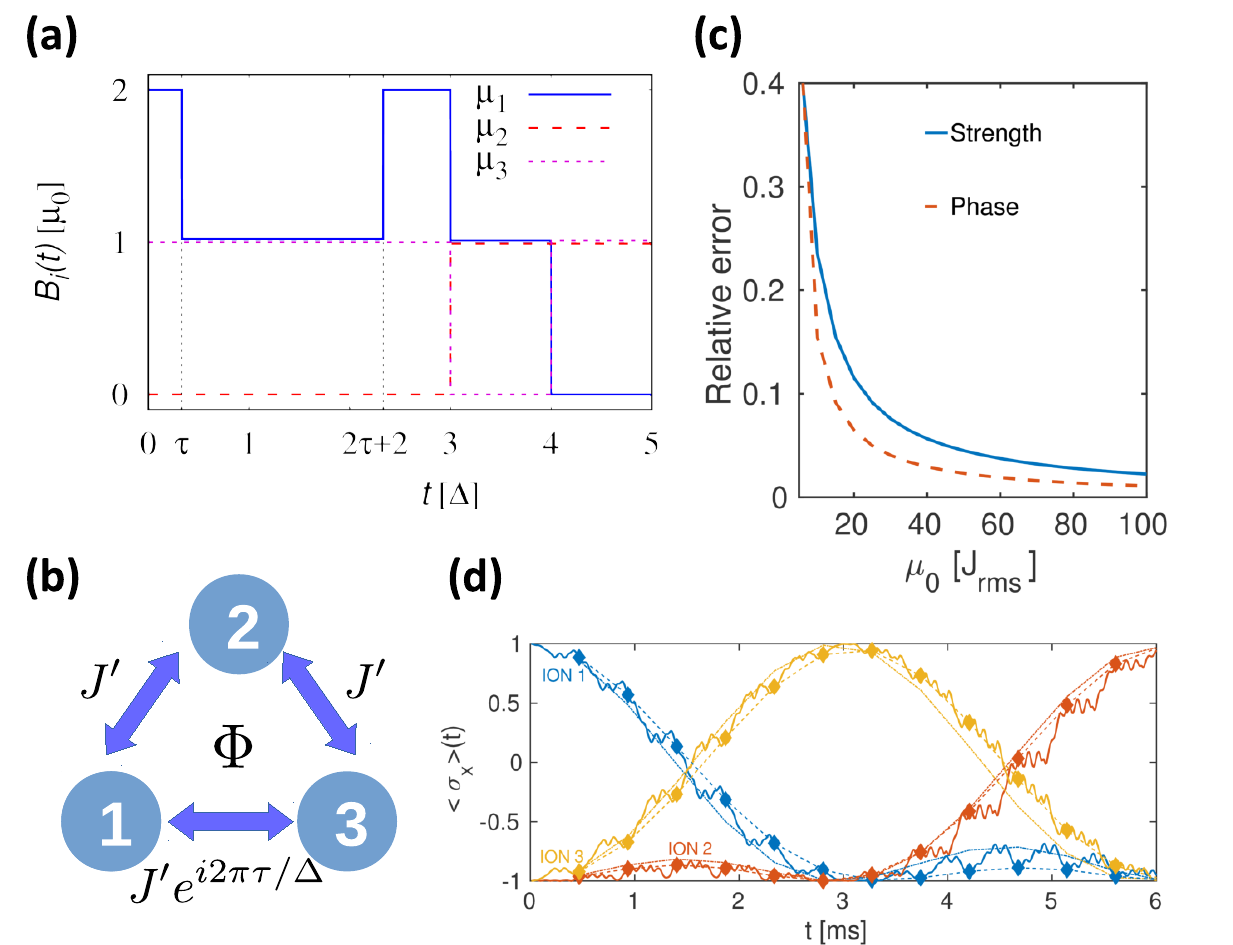}
\caption{\label{protocol} Driving scheme and effective dynamics.
(a) Driving protocol for a chain of three spins: The first sequence ($t<3$) renders interactions between spin 1 (left ion) and 3 (right ion) complex-valued, and suppresses its strength by a factor $2/5$. The sequence for $t>3$ leads to real-valued interactions between nearest neighbors, that is, between spin 1 and 2 (center ion), and between spin 2 and 3. Its strength is $1/5$ of the value without shaking. If nearest neighbor interactions without shaking are engineered twice as strong as second-neighbor interactions via the detuning, the shaking models a situation as depicted in (b): All ions interact equally, encircling a magnetic flux. In (c), the discrepancy between this model and the exact Floquet Hamiltonian, evaluated within a XX model description, is plotted for $\tau/\Delta=1/4$. In (d), we compare the dynamics of the three spins in the Ising model $H_J$ (solid lines) and XX model $H_{XX}$ (dashed-dotted lines). The diamonds (connected by dashed lines as a guide to the eye) mark the stroboscopic evolution at times $t=mT$.
Here, as in rest of the paper, the system parameters are $\delta_{\rm COM}=2\pi\times80\mbox{ kHz}$ for the blue-detuning from the center-of-mass mode, mass $M=171${\rm amu}, trap frequencies $\omega_{XY}=2\pi\times 5 \mbox{ MHz}$ and $\omega_Z=2\pi\times 900 \mbox{ kHz}$. The Rabi frequency is $\Omega=2\pi\times 200\,{\rm kHz}$, recoil frequency is $\omega_{\rm rec}=2\pi\times 26 { \rm\, kHz}$. This choice leads to a root mean square of the spin-spin interactions $J_{\rm rms}=2\pi\times 270{\mbox{ Hz}}$. Field strengths are $B_0=J_{\rm rms}$ and $\mu_0=20 J_{\rm rms}$.
}
\end{figure}

We now apply this general formalism to a concrete driving scheme, similar to the one proposed in Ref. \cite{grass15}. Each potential $\mu_i(t)$ takes piecewise constant values, being integer multiples of some frequency $\mu_0=\pi/\Delta$. Here, $\Delta$ provides the time unit, with $T$ being an integer multiple of $\Delta$.
This choice is motivated in the following way:
\begin{enumerate}
 \item 
 If two potentials remain constant during an interval $\Delta$, the interaction $J'$ between the corresponding spins remains unchanged if the two potentials are the same, or is fully suppressed if the potentials are different, $J'=\frac{J}{\Delta} \int_0^\Delta {\rm e}^{2i m \mu_0 t} {\rm d}t  = J \delta_{m,0}$. This feature allows for engineering the effective strength of interactions.

\item 
 If a potential changes within an interval $\Delta$, this gives rise to complex interaction parameters. For concreteness, let us assume that two potentials differ by $m \mu_0$ for a time $\tau=\Delta/q$, with $m$ an integer. The potentials shall then drop to zero simultaneously for an interval of length $\Delta$, and finally return to their original values for an interval of length $\Delta-\tau$. For such sequence the first order Floquet analysis yields
  \begin{align}
J'= \frac{J}{2\Delta} \int_\tau^{\tau+\Delta} {\rm e}^{2\pi i m/q }{\rm d}t = \frac{J}{2} {\rm e}^{2\pi i m/q }.
\end{align}
This feature allows for generating complex phases, needed to simulate artificial magnetic flux.
\end{enumerate}

According to the reasoning explained above, the shaking period shown in Fig.~\ref{protocol}(a) should (i) generate a complex phase $\varphi/(2\pi) =  -\tau/\Delta$ on the link between spin 1 and 3, and (ii) enhance the ratio $|J_{13}'|/|J_{12}'|=J_{13}'|/|J_{23}'|$ by a factor 2 compared to the original ratio $|J_{13}|/|J_{12}|$. This can fully compensate the decay of interaction strength with distance, and lead to approximately equal interactions between all pairs, cf.  Fig.~\ref{protocol}(b).

The reasoning so far was based on the assumption that time-averaging in the interaction picture approximates well the effective Hamiltonian. Rigorously, though, the effective 
Hamiltonian is obtained from Floquet's theorem. Due to time-periodicity of the evolution operator $U(t,t_0) = U(t+T,t_0+T)$, the operator $U(T,0)$ fully determines the evolution at stroboscopic times $t=mT$. Writing $U(T,0)=\exp\left(-\frac{i}{\hbar} H_{\rm eff} T\right)$, we obtain an effective Hamiltonian $H_{\rm eff}$, which exactly describes the stroboscopic dynamics. For our protocol, consisting only of quenches between time-constant Hamiltonians, $H_{\rm eff}$ can straightforwardly be evaluated. As seen in Fig. \ref{protocol}(c), the discrepancy between the couplings in the exact $H_{\rm eff}$ and the couplings approximated according to Eq. (\ref{jeff}), regarding both absolute value and phase, decreases as $\sim 1/\mu_0$ with the shaking strength. Relative errors $<0.1$ require a shaking $\mu_0 \approx 20 J_{\rm rms}$, where $J_{\rm rms}$ denotes the root mean square of the spin-spin interactions before shaking. We also find that, in order to establish an approximate XX symmetry, a homogeneous $B_0=J_{\rm rms}$ is sufficient. 

Fig. \ref{protocol}(d) compares the spin dynamics in the Ising and the XX model in the presence of a flux $\Phi=\pi/2$ ($\tau=\Delta/4$) after initially preparing the system in a state $\ket{\uparrow \downarrow \downarrow}$ along $\sigma_x$. The curves show good quantitative agreement. Small wiggles which are only seen in the dynamics of the Ising model are due to the strong transverse fields, but do not appear in the stroboscopic evolution. Both models exhibit clear evidence of a chiral current, as the up-spin moves counter-clockwise from ion 1, to ion 3, and finally to ion 2.  Notably, the evolution is almost periodic with a period $T'$, so comparing the states at time $t$ and $T'-t$ allows for a practical detection of time-reversal symmetry breaking \cite{roushan17}.

\begin{figure}[t]
\centering
\includegraphics[width=0.48\textwidth, angle=0]{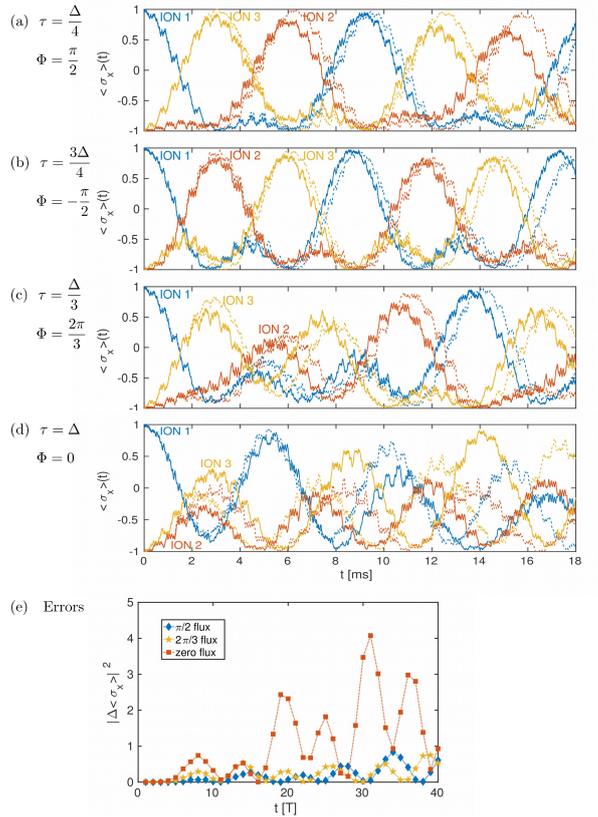}
\caption{\label{dicksim} Comparison between the effective spin dynamics and the full dynamics that includes the phonon modes.
 We plot the spin evolution in the Dicke model (solid lines) and the Ising model (dashed lines) for different parameters $\tau$, adjusting fluxes $\Phi=2\pi\tau/\Delta$. For $\pm\frac{\pi}{2}$-fluxes (a,b), chiral current are observed with good agreement between the two models. For other values of flux, yet different from zero or $\pi$, the two models still produce qualitative agreement, see (c). Without flux, see (d), the Dicke simulation diverges quickly from the Ising model description. In (e), for different values of the flux and at stroboscopic times $t=mT$, the deviations between Ising and Dicke dynamics is plotted, defined as $|\Delta \langle \sigma_x \rangle|^2 \equiv \sum_i \left( \langle \sigma_x^i \rangle_{\rm Ising} - \langle \sigma_x^i \rangle_{\rm Dicke} \right)^2$.
}
\end{figure}

These results establish that Floquet engineering works sufficiently well on the level of an effective Ising system, if shaking strength $\mu_0\sim10 \mbox{ kHz}$ is at least an order of magnitude faster than the spin interactions ($J\sim1 \mbox{ kHz}$ before shaking, $J'=J/5\sim 200\mbox{ Hz}$ after shaking). However, the validity of the spin model description also requires the detuning of the spin-phonon coupling to be fast. A realistic choice is $\delta\sim 100\mbox{ kHz}$, i.e. one order of magnitude above $\mu_0$. To explicitly check the role played by phonons we have simulated the dynamics under the Dicke Hamiltonian $H(t)=H_0(t)+H_B(t)$ using Krylov methods.
The results are seen in Fig. \ref{dicksim} for different fluxes. We first note that the evolution for $\Phi=\pm \frac{\pi}{2}$, shown in panels (a,b), exhibits time-reversal symmetry breaking in the same way as discussed before for the pure spin model. As expected, the direction of the spin current depends on the sign of the flux. For other fractional values of the flux, such as $\Phi=2\pi/3$ shown in panel (c), the evolution is not periodic on relevant time scales, but breaking of time-reversal symmetry can still be infered from obvious differences between clockwise and counter-clockwise flow. In contrast, for ``integer'' fluxes (i.e. ${\rm mod}(\Phi,\pi)=0$) shown in panel (d), the spin flows, at least initially, simultaneously clockwise and counter-clockwise at equal rate. 

Strikingly, time-reversal symmetry breaking seems to stabilize the quantum simulation. For fractional fluxes [panel (a--c)], the evolution in the Dicke model agrees well with the evolution in the Ising model. In contrast to this, the spin dynamics in the system with integer fluxes [panel (d)] is heavily perturbed by the phonons, although only a single parameter $\tau$ has been changed, see also panel (e) plotting the error vs. time. This effect can not be traced to the number of phonons which is approximately the same for different values of $\tau$. We also verified that the effect is visible for different shaking protocols expected to lead to the same effective dynamics. For instance, it appears even stronger if the complex phase is engineered between nearest neighbors rather than second neighbors. While we have no microscopic understanding of the origin of this effect, time-reversal symmetry breaking seems to provide a protection even in our minimal example involving only three ions. We have also carried out analog simulations for a system of four ions, forming two triangles glued together. Differently from the three ions setting, in this case the spin flip hops on more neighbouring sites at once
but the dynamics still exhibits clear signs of chiral flow in the presence of a flux. 
As for three ions, the flux reduces the deviations between spin model dynamics and Dicke model dynamics, although this effect is found to be less pronounced for four ions.

\begin{figure}[t]
\centering
\includegraphics[width=0.48\textwidth, angle=0]{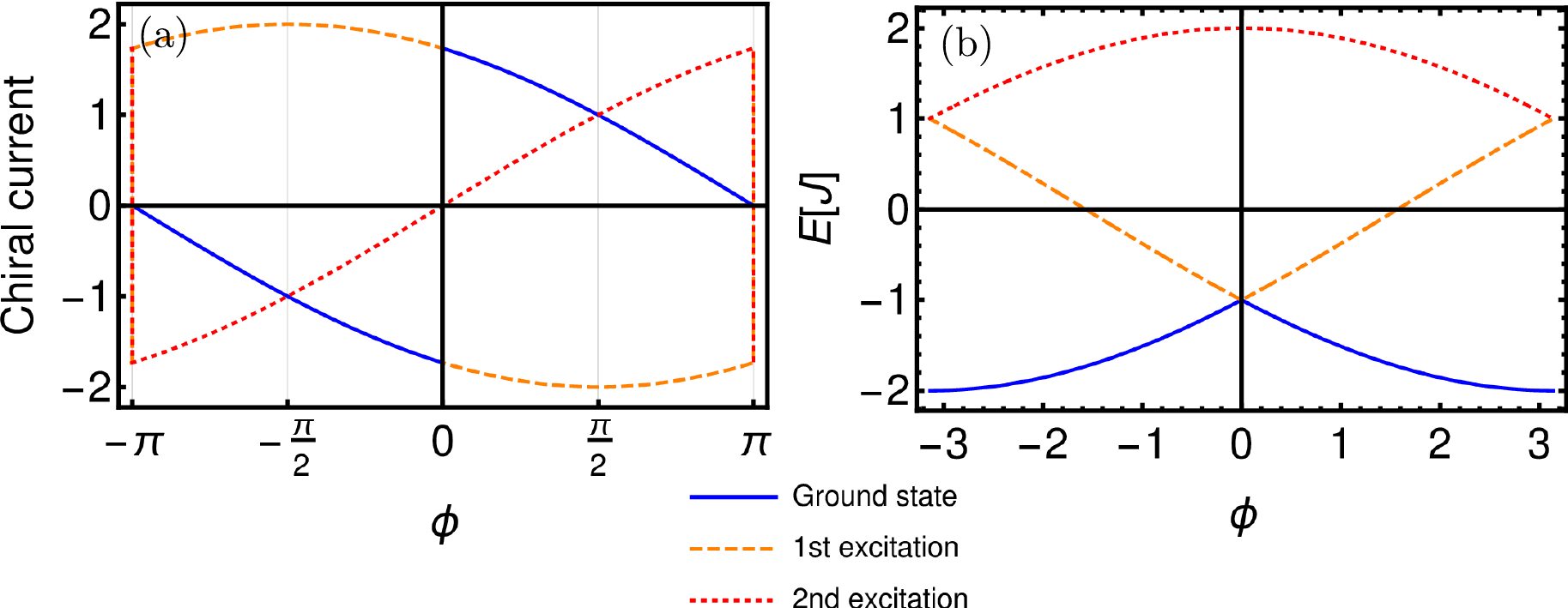}
\caption{\label{triangle} Chiral currents in the XX model on a triangle. We evaluate the equilibrium chiral currents as a function of the flux (a), and the eigenenergies (b). 
}
\end{figure}

So far, we have studied the dynamics seen when the system is initially prepared out of equilibrium. Quantum revivals of the initial state have allowed for demonstrating time-reversal symmetry breaking. However, to keep revival periods sufficiently short, the eigenenergies of states contributing to the initial configuration have to be commensurate. 
While this will always be the case for a triangle at (half-)integer flux [see the energy spectrum as a function of the flux in Fig. \ref{triangle}(b)], revival periods might get long for other fluxes [cf. $2\pi/3$-flux in Fig. \ref{dicksim}(c)] or in larger systems. In such cases, a more general criterion for time-reversal symmetry breaking and chirality might be required. One possibility is to consider equilibrium chiral currents, quantified as \cite{roushan17}
\begin{align}
 I_{\rm chiral} = i \bra{\Psi} \sum_{i\neq j} J_{ij} \sigma^+_i \sigma^-_j - {\rm H.c.} \ket{\Psi},
\end{align}
for an eigenstate $\ket{\Psi}$. As seen in Fig. \ref{triangle}(a), equilibrium chiral currents are non-zero for any eigenstate, if the flux is not an integer of $\pi$. At integer fluxes degenerate eigenstates exist, whose chiral currents add up to zero, and also the current of the non-degenerate eigenstate is zero.

In our previous considerations, the chirality of the spin current served as a direct hallmark of the flux. However, this can barely be used to determine the flux also quantitatively.
In principle, though, trapped ions allow for state tomography providing the full density matrix of the system. This data could then be used to reveal quantitatively the value of the flux felt by the ions. We will now discuss how, for a slightly simplified configuration, information about a complex phase can be extracted already by measuring a single spin-spin correlation function.

\begin{figure}[t]
\centering
\includegraphics[width=0.48\textwidth, angle=0]{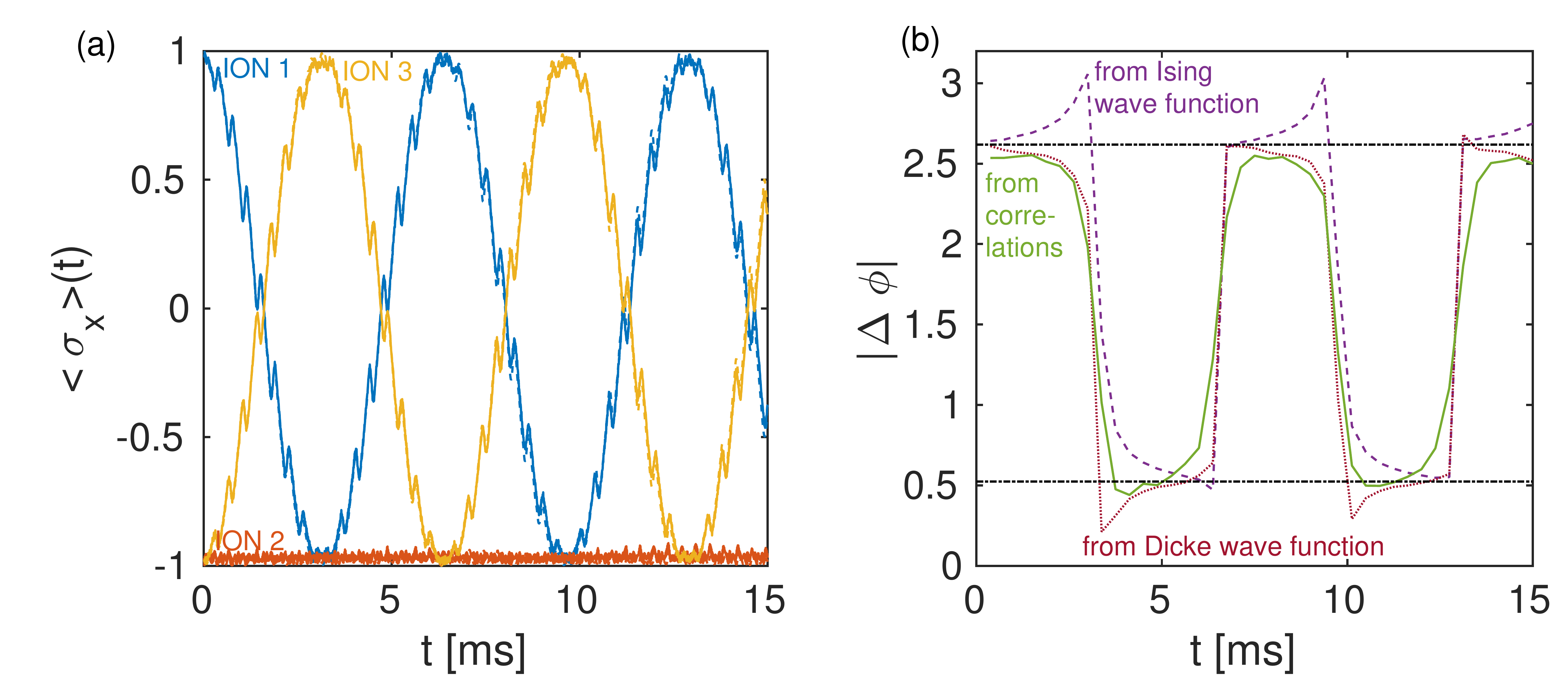}
\caption{\label{rabi} Simplified spin dynamics. (a) Spin evolution for a shaking ($\mu_0=10 J_{\rm rms}$) on the first and the third ion, leaving the central ion detuned at any time. The solid lines (for Dicke model description) almost fully cover the dashed lines (Ising model description). (b) The absolute value of the relative phase between the two states as a function of time, as measured from the wave function in the Ising model, the wave function in the Dicke model, and the correlation functions in the Dicke model. It is defined by the time of the first potential drop, here $\tau=\Delta/3$. The expected phase difference is marked by the black dash-dotted lines.
}
\end{figure}

Therefore, we consider a shaking protocol of period $T=2\Delta$, with $\mu_2=2\mu_0$ always detuned from $\mu_1$ and $\mu_3$ which are both set to zero for $\tau<t\leq\tau+\Delta$, while taking values zero and $\mu_0$ at other times. This protocol leads to particularly simple dynamics, as it freezes out the center spin. Thus, with $\ket{{\rm L}}$ ($\ket{\rm R}$) denoting the position of the up-spin on the left (right) end, we obtain a two-level or double-well system described by  $ H_2 = 2\hbar J_{13}' \ket{\rm L}\bra{\rm R} + {\rm H.c.}$
Accordingly, as seen in Fig. \ref{rabi}(a), the spin dynamics in this case reduces to Rabi oscillations between the left and the right spin with a period $T_{\rm osc} = \pi/|J_{13}'|$. The complex phase $\varphi$ of $J_{13}'$ is reflected in the relative phase between the two levels. An initial state $\Psi(0)=\ket{\rm L}$ will evolve to $ \Psi(t) = \cos(2|J_{13}|'t) \ket{\rm L } + \exp\left[-i\left( \varphi-\frac{\pi}{2} \right) \right] \sin(|J_{13}|'t) \ket{ \rm R}$, where $\varphi$ is the complex phase of $J_{13}'$. Thus, information about the phase is contained in the phase difference $\Delta \phi$ between the two components of the wave function. During the course of time,  $\Delta \phi$ is expected to jump between constant values $\Delta \phi_0= -\frac{\pi}{2}+\varphi = -\pi \left( \frac{1}{2}+ \frac{2\tau}{\Delta} \right)$, when $\sin(|J_{13}|'t)$ and $\cos(|J_{13}|'t)$ have equal signs, and  $\Delta \phi_\pm= \Delta \phi_0 \pm \pi$, if signs are opposite.

The relative phase $\Delta \phi$ can be extracted (up to a sign) from a $\sigma_1^x \sigma_3^x$ measurement. Denoting $\Psi=c_{\rm L} \ket{\rm L} + c_{\rm R}\ket{\rm R}$ we have
\begin{align}
|\Delta \phi| &= {\rm arccos} \left(\frac{\langle \sigma_1^x \sigma_3^x \rangle }{2 |c_{\rm L}| \cdot |c_{\rm R}|} \right), \label{phasemeasure}
\end{align}
where $|c_{\rm L}| = \sqrt{\frac{1}{2} \left( 1+\langle \sigma_1^z \rangle \right)}$, and 
$|c_{\rm R}| = \sqrt{\frac{1}{2} \left( 1-\langle \sigma_1^z \rangle \right)}$.

In Fig. \ref{rabi}(b), we plot $\Delta \phi$, as measured in three different ways: (i) from the wave function in the Ising model, (ii) from the wave function in the Dicke model, (iii) from the correlation functions in the Dicke model according to Eq. (\ref{phasemeasure}). It is seen that all three measures reflect the $\pi$ jumps related to the Rabi oscillation, and are quantitatively close to the expected values ($|\varphi|=5\pi/6$ and $\pi/6$ for $\tau=\Delta/3$). It should be noted, though, that in contrast to the flux through the triangle, the relative phase in the double-well is not a gauge-invariant quantity, and is fully defined within the first shaking period. That is, shifting the initial time within the shaking period, would lead to a different phase.

In summary, we have simulated a system of three ions which encircle an artificial flux, engineered by periodic driving.  Our simulation not only has shown the feasibility of Floquet engineering in trapped ions, but also revealed an unexpected robustness of the spin dynamics when the driving breaks time-reversal symmetry in the effective model. This effect could stabilize and further enhance the quantum simulation of topological models also in larger systems. Having shown a path towards complex-valued spin-spin interaction, we believe that trapped ions provide an ideal system for studying the role of topology and topological protection in small clusters. Potentially, such clusters will provide robust building blocks for larger quantum devices.

\textit{ Acknowledgements.}
We are grateful to Alexey Gorshkov, Bruno Jul{\'i}a-D{\'i}az, Jiehang Zhang, Paul Hess, and Zhexuan Gong for interesting discussions.
Financial support from AFOSR-MURI, EU grants EQuaM (FP7/2007-2013 Grant
No. 323714), OSYRIS (ERC-2013-AdG Grant No. 339106), SIQS (FP7-ICT-2011-9 No. 600645), QUIC (H2020-FETPROACT-2014 No. 641122), Spanish MINECO (SEVERO OCHOA Grant SEV-2015-0522, FISICATEAMO FIS2016-79508-P), the Generalitat de Catalunya (SGR 874 and CERCA program), Fundacio Privada Cellex is acknowledged. GP is supported by the IC postdoctoral Research Fellowship program.


\begin{thebibliography}{50}%
\makeatletter
\providecommand \@ifxundefined [1]{%
 \@ifx{#1\undefined}
}%
\providecommand \@ifnum [1]{%
 \ifnum #1\expandafter \@firstoftwo
 \else \expandafter \@secondoftwo
 \fi
}%
\providecommand \@ifx [1]{%
 \ifx #1\expandafter \@firstoftwo
 \else \expandafter \@secondoftwo
 \fi
}%
\providecommand \natexlab [1]{#1}%
\providecommand \enquote  [1]{``#1''}%
\providecommand \bibnamefont  [1]{#1}%
\providecommand \bibfnamefont [1]{#1}%
\providecommand \citenamefont [1]{#1}%
\providecommand \href@noop [0]{\@secondoftwo}%
\providecommand \href [0]{\begingroup \@sanitize@url \@href}%
\providecommand \@href[1]{\@@startlink{#1}\@@href}%
\providecommand \@@href[1]{\endgroup#1\@@endlink}%
\providecommand \@sanitize@url [0]{\catcode `\\12\catcode `\$12\catcode
  `\&12\catcode `\#12\catcode `\^12\catcode `\_12\catcode `\%12\relax}%
\providecommand \@@startlink[1]{}%
\providecommand \@@endlink[0]{}%
\providecommand \url  [0]{\begingroup\@sanitize@url \@url }%
\providecommand \@url [1]{\endgroup\@href {#1}{\urlprefix }}%
\providecommand \urlprefix  [0]{URL }%
\providecommand \Eprint [0]{\href }%
\providecommand \doibase [0]{http://dx.doi.org/}%
\providecommand \selectlanguage [0]{\@gobble}%
\providecommand \bibinfo  [0]{\@secondoftwo}%
\providecommand \bibfield  [0]{\@secondoftwo}%
\providecommand \translation [1]{[#1]}%
\providecommand \BibitemOpen [0]{}%
\providecommand \bibitemStop [0]{}%
\providecommand \bibitemNoStop [0]{.\EOS\space}%
\providecommand \EOS [0]{\spacefactor3000\relax}%
\providecommand \BibitemShut  [1]{\csname bibitem#1\endcsname}%
\let\auto@bib@innerbib\@empty
\bibitem [{\citenamefont {Aharonov}\ and\ \citenamefont {Bohm}(1959)}]{bohm}%
  \BibitemOpen
  \bibfield  {author} {\bibinfo {author} {\bibfnamefont {Y.}~\bibnamefont
  {Aharonov}}\ and\ \bibinfo {author} {\bibfnamefont {D.}~\bibnamefont
  {Bohm}},\ }\href {\doibase 10.1103/PhysRev.115.485} {\bibfield  {journal}
  {\bibinfo  {journal} {Phys. Rev.}\ }\textbf {\bibinfo {volume} {115}},\
  \bibinfo {pages} {485} (\bibinfo {year} {1959})}\BibitemShut {NoStop}%
\bibitem [{\citenamefont {Shapere}\ and\ \citenamefont
  {Wilczek}(1989)}]{Wilczek-book}%
  \BibitemOpen
  \bibfield  {author} {\bibinfo {author} {\bibfnamefont {A.}~\bibnamefont
  {Shapere}}\ and\ \bibinfo {author} {\bibfnamefont {F.}~\bibnamefont
  {Wilczek}},\ }\href@noop {} {\emph {\bibinfo {title} {{Geometric Phases in
  Physics}}}},\ {Advanced Series in Mathematical Physics Vol. 5}\ (\bibinfo
  {publisher} {World Scientific},\ \bibinfo {year} {1989})\BibitemShut
  {NoStop}%
\bibitem [{\citenamefont {Hofstadter}(1976)}]{hofstadter}%
  \BibitemOpen
  \bibfield  {author} {\bibinfo {author} {\bibfnamefont {D.~R.}\ \bibnamefont
  {Hofstadter}},\ }\href {\doibase 10.1103/PhysRevB.14.2239} {\bibfield
  {journal} {\bibinfo  {journal} {Phys. Rev. B}\ }\textbf {\bibinfo {volume}
  {14}},\ \bibinfo {pages} {2239} (\bibinfo {year} {1976})}\BibitemShut
  {NoStop}%
\bibitem [{\citenamefont {Janssen}\ \emph {et~al.}(1994)\citenamefont
  {Janssen}, \citenamefont {Viehweger}, \citenamefont {Fastenrath},\ and\
  \citenamefont {Hajdu}}]{IQHE-book}%
  \BibitemOpen
  \bibfield  {author} {\bibinfo {author} {\bibfnamefont {M.}~\bibnamefont
  {Janssen}}, \bibinfo {author} {\bibfnamefont {O.}~\bibnamefont {Viehweger}},
  \bibinfo {author} {\bibfnamefont {U.}~\bibnamefont {Fastenrath}}, \ and\
  \bibinfo {author} {\bibfnamefont {J.}~\bibnamefont {Hajdu}},\ }\href@noop {}
  {\emph {\bibinfo {title} {{Introduction to the Theory of the Integer Qunantum
  Hall Effect}}}}\ (\bibinfo  {publisher} {Wiley, John and Sons},\ \bibinfo
  {year} {1994})\BibitemShut {NoStop}%
\bibitem [{\citenamefont {Laughlin}(1983)}]{laughlin}%
  \BibitemOpen
  \bibfield  {author} {\bibinfo {author} {\bibfnamefont {R.~B.}\ \bibnamefont
  {Laughlin}},\ }\href {\doibase 10.1103/PhysRevLett.50.1395} {\bibfield
  {journal} {\bibinfo  {journal} {Phys. Rev. Lett.}\ }\textbf {\bibinfo
  {volume} {50}},\ \bibinfo {pages} {1395} (\bibinfo {year}
  {1983})}\BibitemShut {NoStop}%
\bibitem [{\citenamefont {Wen}(2004)}]{Wen-book}%
  \BibitemOpen
  \bibfield  {author} {\bibinfo {author} {\bibfnamefont {X.}~\bibnamefont
  {Wen}},\ }\href {https://books.google.com/books?id=llnlrfdR4YgC} {\emph
  {\bibinfo {title} {{Quantum Field Theory of Many-Body Systems}}}},\ {Oxford
  Graduate Texts}\ (\bibinfo  {publisher} {OUP Oxford},\ \bibinfo {year}
  {2004})\BibitemShut {NoStop}%
\bibitem [{\citenamefont {Nayak}\ \emph {et~al.}(2008)\citenamefont {Nayak},
  \citenamefont {Simon}, \citenamefont {Stern}, \citenamefont {Freedman},\ and\
  \citenamefont {{Das Sarma}}}]{nayak}%
  \BibitemOpen
  \bibfield  {author} {\bibinfo {author} {\bibfnamefont {C.}~\bibnamefont
  {Nayak}}, \bibinfo {author} {\bibfnamefont {S.~H.}\ \bibnamefont {Simon}},
  \bibinfo {author} {\bibfnamefont {A.}~\bibnamefont {Stern}}, \bibinfo
  {author} {\bibfnamefont {M.}~\bibnamefont {Freedman}}, \ and\ \bibinfo
  {author} {\bibfnamefont {S.}~\bibnamefont {{Das Sarma}}},\ }\href {\doibase
  10.1103/RevModPhys.80.1083} {\bibfield  {journal} {\bibinfo  {journal} {Rev.
  Mod. Phys.}\ }\textbf {\bibinfo {volume} {80}},\ \bibinfo {pages} {1083}
  (\bibinfo {year} {2008})}\BibitemShut {NoStop}%
\bibitem [{\citenamefont {Struck}\ \emph {et~al.}(2012)\citenamefont {Struck},
  \citenamefont {{\"O}lschl{\"a}ger}, \citenamefont {Weinberg}, \citenamefont
  {Hauke}, \citenamefont {Simonet}, \citenamefont {Eckardt}, \citenamefont
  {Lewenstein}, \citenamefont {Sengstock},\ and\ \citenamefont
  {Windpassinger}}]{struck12}%
  \BibitemOpen
  \bibfield  {author} {\bibinfo {author} {\bibfnamefont {J.}~\bibnamefont
  {Struck}}, \bibinfo {author} {\bibfnamefont {C.}~\bibnamefont
  {{\"O}lschl{\"a}ger}}, \bibinfo {author} {\bibfnamefont {M.}~\bibnamefont
  {Weinberg}}, \bibinfo {author} {\bibfnamefont {P.}~\bibnamefont {Hauke}},
  \bibinfo {author} {\bibfnamefont {J.}~\bibnamefont {Simonet}}, \bibinfo
  {author} {\bibfnamefont {A.}~\bibnamefont {Eckardt}}, \bibinfo {author}
  {\bibfnamefont {M.}~\bibnamefont {Lewenstein}}, \bibinfo {author}
  {\bibfnamefont {K.}~\bibnamefont {Sengstock}}, \ and\ \bibinfo {author}
  {\bibfnamefont {P.}~\bibnamefont {Windpassinger}},\ }\href {\doibase
  10.1103/PhysRevLett.108.225304} {\bibfield  {journal} {\bibinfo  {journal}
  {Phys. Rev. Lett.}\ }\textbf {\bibinfo {volume} {108}},\ \bibinfo {pages}
  {225304} (\bibinfo {year} {2012})}\BibitemShut {NoStop}%
\bibitem [{\citenamefont {Miyake}\ \emph {et~al.}(2013)\citenamefont {Miyake},
  \citenamefont {Siviloglou}, \citenamefont {Kennedy}, \citenamefont {Burton},\
  and\ \citenamefont {Ketterle}}]{miyake13}%
  \BibitemOpen
  \bibfield  {author} {\bibinfo {author} {\bibfnamefont {H.}~\bibnamefont
  {Miyake}}, \bibinfo {author} {\bibfnamefont {G.~A.}\ \bibnamefont
  {Siviloglou}}, \bibinfo {author} {\bibfnamefont {C.~J.}\ \bibnamefont
  {Kennedy}}, \bibinfo {author} {\bibfnamefont {W.~C.}\ \bibnamefont {Burton}},
  \ and\ \bibinfo {author} {\bibfnamefont {W.}~\bibnamefont {Ketterle}},\
  }\href {\doibase 10.1103/PhysRevLett.111.185302} {\bibfield  {journal}
  {\bibinfo  {journal} {Phys. Rev. Lett.}\ }\textbf {\bibinfo {volume} {111}},\
  \bibinfo {pages} {185302} (\bibinfo {year} {2013})}\BibitemShut {NoStop}%
\bibitem [{\citenamefont {Aidelsburger}\ \emph {et~al.}(2013)\citenamefont
  {Aidelsburger}, \citenamefont {Atala}, \citenamefont {Lohse}, \citenamefont
  {Barreiro}, \citenamefont {Paredes},\ and\ \citenamefont
  {Bloch}}]{aidelsburger13}%
  \BibitemOpen
  \bibfield  {author} {\bibinfo {author} {\bibfnamefont {M.}~\bibnamefont
  {Aidelsburger}}, \bibinfo {author} {\bibfnamefont {M.}~\bibnamefont {Atala}},
  \bibinfo {author} {\bibfnamefont {M.}~\bibnamefont {Lohse}}, \bibinfo
  {author} {\bibfnamefont {J.~T.}\ \bibnamefont {Barreiro}}, \bibinfo {author}
  {\bibfnamefont {B.}~\bibnamefont {Paredes}}, \ and\ \bibinfo {author}
  {\bibfnamefont {I.}~\bibnamefont {Bloch}},\ }\href {\doibase
  10.1103/PhysRevLett.111.185301} {\bibfield  {journal} {\bibinfo  {journal}
  {Phys. Rev. Lett.}\ }\textbf {\bibinfo {volume} {111}},\ \bibinfo {pages}
  {185301} (\bibinfo {year} {2013})}\BibitemShut {NoStop}%
\bibitem [{\citenamefont {Jotzu}\ \emph {et~al.}(2014)\citenamefont {Jotzu},
  \citenamefont {Messer}, \citenamefont {Desbuquois}, \citenamefont {Lebrat},
  \citenamefont {Uehlinger}, \citenamefont {Greif},\ and\ \citenamefont
  {Esslinger}}]{jotzu14}%
  \BibitemOpen
  \bibfield  {author} {\bibinfo {author} {\bibfnamefont {G.}~\bibnamefont
  {Jotzu}}, \bibinfo {author} {\bibfnamefont {M.}~\bibnamefont {Messer}},
  \bibinfo {author} {\bibfnamefont {R.}~\bibnamefont {Desbuquois}}, \bibinfo
  {author} {\bibfnamefont {M.}~\bibnamefont {Lebrat}}, \bibinfo {author}
  {\bibfnamefont {T.}~\bibnamefont {Uehlinger}}, \bibinfo {author}
  {\bibfnamefont {D.}~\bibnamefont {Greif}}, \ and\ \bibinfo {author}
  {\bibfnamefont {T.}~\bibnamefont {Esslinger}},\ }\href {\doibase
  10.1038/nature13915;;;;;;;; 10.1038/nature13915} {\bibfield  {journal}
  {\bibinfo  {journal} {Nature}\ }\textbf {\bibinfo {volume} {515}},\ \bibinfo
  {pages} {237} (\bibinfo {year} {2014})}\BibitemShut {NoStop}%
\bibitem [{\citenamefont {Mancini}\ \emph {et~al.}(2015)\citenamefont
  {Mancini}, \citenamefont {Pagano}, \citenamefont {Cappellini}, \citenamefont
  {Livi}, \citenamefont {Rider}, \citenamefont {Catani}, \citenamefont {Sias},
  \citenamefont {Zoller}, \citenamefont {Inguscio}, \citenamefont {Dalmonte},\
  and\ \citenamefont {Fallani}}]{Mancini15}%
  \BibitemOpen
  \bibfield  {author} {\bibinfo {author} {\bibfnamefont {M.}~\bibnamefont
  {Mancini}}, \bibinfo {author} {\bibfnamefont {G.}~\bibnamefont {Pagano}},
  \bibinfo {author} {\bibfnamefont {G.}~\bibnamefont {Cappellini}}, \bibinfo
  {author} {\bibfnamefont {L.}~\bibnamefont {Livi}}, \bibinfo {author}
  {\bibfnamefont {M.}~\bibnamefont {Rider}}, \bibinfo {author} {\bibfnamefont
  {J.}~\bibnamefont {Catani}}, \bibinfo {author} {\bibfnamefont
  {C.}~\bibnamefont {Sias}}, \bibinfo {author} {\bibfnamefont {P.}~\bibnamefont
  {Zoller}}, \bibinfo {author} {\bibfnamefont {M.}~\bibnamefont {Inguscio}},
  \bibinfo {author} {\bibfnamefont {M.}~\bibnamefont {Dalmonte}}, \ and\
  \bibinfo {author} {\bibfnamefont {L.}~\bibnamefont {Fallani}},\ }\href
  {\doibase 10.1126/science.aaa8736} {\bibfield  {journal} {\bibinfo  {journal}
  {Science}\ }\textbf {\bibinfo {volume} {349}},\ \bibinfo {pages} {1510}
  (\bibinfo {year} {2015})}\BibitemShut {NoStop}%
\bibitem [{\citenamefont {Stuhl}\ \emph {et~al.}(2015)\citenamefont {Stuhl},
  \citenamefont {Lu}, \citenamefont {Aycock}, \citenamefont {Genkina},\ and\
  \citenamefont {Spielman}}]{stuhl15}%
  \BibitemOpen
  \bibfield  {author} {\bibinfo {author} {\bibfnamefont {B.~K.}\ \bibnamefont
  {Stuhl}}, \bibinfo {author} {\bibfnamefont {H.-I.}\ \bibnamefont {Lu}},
  \bibinfo {author} {\bibfnamefont {L.~M.}\ \bibnamefont {Aycock}}, \bibinfo
  {author} {\bibfnamefont {D.}~\bibnamefont {Genkina}}, \ and\ \bibinfo
  {author} {\bibfnamefont {I.~B.}\ \bibnamefont {Spielman}},\ }\href {\doibase
  10.1126/science.aaa8515} {\bibfield  {journal} {\bibinfo  {journal}
  {Science}\ }\textbf {\bibinfo {volume} {349}},\ \bibinfo {pages} {1514}
  (\bibinfo {year} {2015})}\BibitemShut {NoStop}%
\bibitem [{\citenamefont {Meinert}\ \emph {et~al.}(2016)\citenamefont
  {Meinert}, \citenamefont {Mark}, \citenamefont {Lauber}, \citenamefont
  {Daley},\ and\ \citenamefont {N{\"a}gerl}}]{meinert16}%
  \BibitemOpen
  \bibfield  {author} {\bibinfo {author} {\bibfnamefont {F.}~\bibnamefont
  {Meinert}}, \bibinfo {author} {\bibfnamefont {M.~J.}\ \bibnamefont {Mark}},
  \bibinfo {author} {\bibfnamefont {K.}~\bibnamefont {Lauber}}, \bibinfo
  {author} {\bibfnamefont {A.~J.}\ \bibnamefont {Daley}}, \ and\ \bibinfo
  {author} {\bibfnamefont {H.-C.}\ \bibnamefont {N{\"a}gerl}},\ }\href
  {\doibase 10.1103/PhysRevLett.116.205301} {\bibfield  {journal} {\bibinfo
  {journal} {Phys. Rev. Lett.}\ }\textbf {\bibinfo {volume} {116}},\ \bibinfo
  {pages} {205301} (\bibinfo {year} {2016})}\BibitemShut {NoStop}%
\bibitem [{\citenamefont {Hafezi}\ \emph {et~al.}(2013)\citenamefont {Hafezi},
  \citenamefont {Mittal}, \citenamefont {Fan}, \citenamefont {Migdall},\ and\
  \citenamefont {Taylor}}]{hafezi13}%
  \BibitemOpen
  \bibfield  {author} {\bibinfo {author} {\bibfnamefont {M.}~\bibnamefont
  {Hafezi}}, \bibinfo {author} {\bibfnamefont {S.}~\bibnamefont {Mittal}},
  \bibinfo {author} {\bibfnamefont {J.}~\bibnamefont {Fan}}, \bibinfo {author}
  {\bibfnamefont {A.}~\bibnamefont {Migdall}}, \ and\ \bibinfo {author}
  {\bibfnamefont {J.}~\bibnamefont {Taylor}},\ }\href {\doibase
  10.1038/nphoton.2013.274;;;; 10.1038/nphoton.2013.274} {\bibfield  {journal}
  {\bibinfo  {journal} {Nat. Photon.}\ }\textbf {\bibinfo {volume} {7}},\
  \bibinfo {pages} {1001} (\bibinfo {year} {2013})}\BibitemShut {NoStop}%
\bibitem [{\citenamefont {Schine}\ \emph {et~al.}(2016)\citenamefont {Schine},
  \citenamefont {Ryou}, \citenamefont {Gromov}, \citenamefont {Sommer},\ and\
  \citenamefont {Simon}}]{schine16}%
  \BibitemOpen
  \bibfield  {author} {\bibinfo {author} {\bibfnamefont {N.}~\bibnamefont
  {Schine}}, \bibinfo {author} {\bibfnamefont {A.}~\bibnamefont {Ryou}},
  \bibinfo {author} {\bibfnamefont {A.}~\bibnamefont {Gromov}}, \bibinfo
  {author} {\bibfnamefont {A.}~\bibnamefont {Sommer}}, \ and\ \bibinfo {author}
  {\bibfnamefont {J.}~\bibnamefont {Simon}},\ }\href {\doibase
  10.1038/nature17943;;; 10.1038/nature17943} {\bibfield  {journal} {\bibinfo
  {journal} {Nature}\ }\textbf {\bibinfo {volume} {534}},\ \bibinfo {pages}
  {671} (\bibinfo {year} {2016})}\BibitemShut {NoStop}%
\bibitem [{\citenamefont {{P. Roushan \textit{ et al.}}}(2017)}]{roushan17}%
  \BibitemOpen
  \bibfield  {author} {\bibinfo {author} {\bibnamefont {{P. Roushan \textit{ et
  al.}}}},\ }\href {\doibase 10.1038/nphys3930;;;;; 10.1038/nphys3930}
  {\bibfield  {journal} {\bibinfo  {journal} {Nat. Phys.}\ }\textbf {\bibinfo
  {volume} {13}},\ \bibinfo {pages} {146} (\bibinfo {year} {2017})}\BibitemShut
  {NoStop}%
\bibitem [{\citenamefont {Lodahl}\ \emph {et~al.}(2017)\citenamefont {Lodahl},
  \citenamefont {Mahmoodian}, \citenamefont {Stobbe}, \citenamefont
  {Rauschenbeutel}, \citenamefont {Schneeweiss}, \citenamefont {Volz},
  \citenamefont {Pichler},\ and\ \citenamefont {Zoller}}]{review_chiralqo_17}%
  \BibitemOpen
  \bibfield  {author} {\bibinfo {author} {\bibfnamefont {P.}~\bibnamefont
  {Lodahl}}, \bibinfo {author} {\bibfnamefont {S.}~\bibnamefont {Mahmoodian}},
  \bibinfo {author} {\bibfnamefont {S.}~\bibnamefont {Stobbe}}, \bibinfo
  {author} {\bibfnamefont {A.}~\bibnamefont {Rauschenbeutel}}, \bibinfo
  {author} {\bibfnamefont {P.}~\bibnamefont {Schneeweiss}}, \bibinfo {author}
  {\bibfnamefont {J.}~\bibnamefont {Volz}}, \bibinfo {author} {\bibfnamefont
  {H.}~\bibnamefont {Pichler}}, \ and\ \bibinfo {author} {\bibfnamefont
  {P.}~\bibnamefont {Zoller}},\ }\href {\doibase 10.1038/nature21037;;;
  10.1038/nature21037} {\bibfield  {journal} {\bibinfo  {journal} {Nature}\
  }\textbf {\bibinfo {volume} {541}},\ \bibinfo {pages} {473} (\bibinfo {year}
  {2017})}\BibitemShut {NoStop}%
\bibitem [{\citenamefont {Kong}\ \emph {et~al.}(2016)\citenamefont {Kong},
  \citenamefont {Ju}, \citenamefont {Liu}, \citenamefont {Lei}, \citenamefont
  {Wang}, \citenamefont {Kong}, \citenamefont {Wang}, \citenamefont {Huang},
  \citenamefont {Li}, \citenamefont {Shi}, \citenamefont {Jiang},\ and\
  \citenamefont {Du}}]{kong16}%
  \BibitemOpen
  \bibfield  {author} {\bibinfo {author} {\bibfnamefont {F.}~\bibnamefont
  {Kong}}, \bibinfo {author} {\bibfnamefont {C.}~\bibnamefont {Ju}}, \bibinfo
  {author} {\bibfnamefont {Y.}~\bibnamefont {Liu}}, \bibinfo {author}
  {\bibfnamefont {C.}~\bibnamefont {Lei}}, \bibinfo {author} {\bibfnamefont
  {M.}~\bibnamefont {Wang}}, \bibinfo {author} {\bibfnamefont {X.}~\bibnamefont
  {Kong}}, \bibinfo {author} {\bibfnamefont {P.}~\bibnamefont {Wang}}, \bibinfo
  {author} {\bibfnamefont {P.}~\bibnamefont {Huang}}, \bibinfo {author}
  {\bibfnamefont {Z.}~\bibnamefont {Li}}, \bibinfo {author} {\bibfnamefont
  {F.}~\bibnamefont {Shi}}, \bibinfo {author} {\bibfnamefont {L.}~\bibnamefont
  {Jiang}}, \ and\ \bibinfo {author} {\bibfnamefont {J.}~\bibnamefont {Du}},\
  }\href {\doibase 10.1103/PhysRevLett.117.060503} {\bibfield  {journal}
  {\bibinfo  {journal} {Phys. Rev. Lett.}\ }\textbf {\bibinfo {volume} {117}},\
  \bibinfo {pages} {060503} (\bibinfo {year} {2016})}\BibitemShut {NoStop}%
\bibitem [{\citenamefont {Mintert}\ and\ \citenamefont
  {Wunderlich}(2001)}]{mintert01}%
  \BibitemOpen
  \bibfield  {author} {\bibinfo {author} {\bibfnamefont {F.}~\bibnamefont
  {Mintert}}\ and\ \bibinfo {author} {\bibfnamefont {C.}~\bibnamefont
  {Wunderlich}},\ }\href {\doibase 10.1103/PhysRevLett.87.257904} {\bibfield
  {journal} {\bibinfo  {journal} {Phys. Rev. Lett.}\ }\textbf {\bibinfo
  {volume} {87}},\ \bibinfo {pages} {257904} (\bibinfo {year}
  {2001})}\BibitemShut {NoStop}%
\bibitem [{\citenamefont {Porras}\ and\ \citenamefont
  {Cirac}(2004)}]{porras04}%
  \BibitemOpen
  \bibfield  {author} {\bibinfo {author} {\bibfnamefont {D.}~\bibnamefont
  {Porras}}\ and\ \bibinfo {author} {\bibfnamefont {J.~I.}\ \bibnamefont
  {Cirac}},\ }\href {\doibase 10.1103/PhysRevLett.92.207901} {\bibfield
  {journal} {\bibinfo  {journal} {Phys. Rev. Lett.}\ }\textbf {\bibinfo
  {volume} {92}},\ \bibinfo {pages} {207901} (\bibinfo {year}
  {2004})}\BibitemShut {NoStop}%
\bibitem [{\citenamefont {Friedenauer}\ \emph {et~al.}(2008)\citenamefont
  {Friedenauer}, \citenamefont {Schmitz}, \citenamefont {Glueckert},
  \citenamefont {Porras},\ and\ \citenamefont {Schaetz}}]{friedenauer08}%
  \BibitemOpen
  \bibfield  {author} {\bibinfo {author} {\bibfnamefont {A.}~\bibnamefont
  {Friedenauer}}, \bibinfo {author} {\bibfnamefont {H.}~\bibnamefont
  {Schmitz}}, \bibinfo {author} {\bibfnamefont {J.~T.}\ \bibnamefont
  {Glueckert}}, \bibinfo {author} {\bibfnamefont {D.}~\bibnamefont {Porras}}, \
  and\ \bibinfo {author} {\bibfnamefont {T.}~\bibnamefont {Schaetz}},\
  }\href@noop {} {\bibfield  {journal} {\bibinfo  {journal} {Nat. Phys.}\
  }\textbf {\bibinfo {volume} {4}},\ \bibinfo {pages} {757} (\bibinfo {year}
  {2008})}\BibitemShut {NoStop}%
\bibitem [{\citenamefont {Kim}\ \emph {et~al.}(2009)\citenamefont {Kim},
  \citenamefont {Chang}, \citenamefont {Islam}, \citenamefont {Korenblit},
  \citenamefont {Duan},\ and\ \citenamefont {Monroe}}]{kim09}%
  \BibitemOpen
  \bibfield  {author} {\bibinfo {author} {\bibfnamefont {K.}~\bibnamefont
  {Kim}}, \bibinfo {author} {\bibfnamefont {M.-S.}\ \bibnamefont {Chang}},
  \bibinfo {author} {\bibfnamefont {R.}~\bibnamefont {Islam}}, \bibinfo
  {author} {\bibfnamefont {S.}~\bibnamefont {Korenblit}}, \bibinfo {author}
  {\bibfnamefont {L.-M.}\ \bibnamefont {Duan}}, \ and\ \bibinfo {author}
  {\bibfnamefont {C.}~\bibnamefont {Monroe}},\ }\href {\doibase
  10.1103/PhysRevLett.103.120502} {\bibfield  {journal} {\bibinfo  {journal}
  {Phys. Rev. Lett.}\ }\textbf {\bibinfo {volume} {103}},\ \bibinfo {pages}
  {120502} (\bibinfo {year} {2009})}\BibitemShut {NoStop}%
\bibitem [{\citenamefont {Kim}\ \emph {et~al.}(2010)\citenamefont {Kim},
  \citenamefont {Chang}, \citenamefont {Korenblit}, \citenamefont {Islam},
  \citenamefont {Edwards}, \citenamefont {Freericks}, \citenamefont {Lin},
  \citenamefont {Duan},\ and\ \citenamefont {Monroe}}]{kim10}%
  \BibitemOpen
  \bibfield  {author} {\bibinfo {author} {\bibfnamefont {K.}~\bibnamefont
  {Kim}}, \bibinfo {author} {\bibfnamefont {M.~S.}\ \bibnamefont {Chang}},
  \bibinfo {author} {\bibfnamefont {S.}~\bibnamefont {Korenblit}}, \bibinfo
  {author} {\bibfnamefont {R.}~\bibnamefont {Islam}}, \bibinfo {author}
  {\bibfnamefont {E.~E.}\ \bibnamefont {Edwards}}, \bibinfo {author}
  {\bibfnamefont {J.~K.}\ \bibnamefont {Freericks}}, \bibinfo {author}
  {\bibfnamefont {G.~D.}\ \bibnamefont {Lin}}, \bibinfo {author} {\bibfnamefont
  {L.~M.}\ \bibnamefont {Duan}}, \ and\ \bibinfo {author} {\bibfnamefont
  {C.}~\bibnamefont {Monroe}},\ }\href {\doibase 10.1038/nature09071}
  {\bibfield  {journal} {\bibinfo  {journal} {Nature}\ }\textbf {\bibinfo
  {volume} {465}},\ \bibinfo {pages} {590} (\bibinfo {year}
  {2010})}\BibitemShut {NoStop}%
\bibitem [{\citenamefont {Richerme}\ \emph {et~al.}(2014)\citenamefont
  {Richerme}, \citenamefont {Gong}, \citenamefont {Lee}, \citenamefont {Senko},
  \citenamefont {Smith}, \citenamefont {Foss-Feig}, \citenamefont {Michalakis},
  \citenamefont {Gorshkov},\ and\ \citenamefont {Monroe}}]{richerme14}%
  \BibitemOpen
  \bibfield  {author} {\bibinfo {author} {\bibfnamefont {P.}~\bibnamefont
  {Richerme}}, \bibinfo {author} {\bibfnamefont {Z.-X.}\ \bibnamefont {Gong}},
  \bibinfo {author} {\bibfnamefont {A.}~\bibnamefont {Lee}}, \bibinfo {author}
  {\bibfnamefont {C.}~\bibnamefont {Senko}}, \bibinfo {author} {\bibfnamefont
  {J.}~\bibnamefont {Smith}}, \bibinfo {author} {\bibfnamefont
  {M.}~\bibnamefont {Foss-Feig}}, \bibinfo {author} {\bibfnamefont
  {S.}~\bibnamefont {Michalakis}}, \bibinfo {author} {\bibfnamefont {A.~V.}\
  \bibnamefont {Gorshkov}}, \ and\ \bibinfo {author} {\bibfnamefont
  {C.}~\bibnamefont {Monroe}},\ }\href {\doibase 10.1038/nature13450;;;;;;;;
  10.1038/nature13450} {\bibfield  {journal} {\bibinfo  {journal} {Nature}\
  }\textbf {\bibinfo {volume} {511}},\ \bibinfo {pages} {198} (\bibinfo {year}
  {2014})}\BibitemShut {NoStop}%
\bibitem [{\citenamefont {Jurcevic}\ \emph {et~al.}(2014)\citenamefont
  {Jurcevic}, \citenamefont {Lanyon}, \citenamefont {Hauke}, \citenamefont
  {Hempel}, \citenamefont {Zoller}, \citenamefont {Blatt},\ and\ \citenamefont
  {Roos}}]{jurcevic14}%
  \BibitemOpen
  \bibfield  {author} {\bibinfo {author} {\bibfnamefont {P.}~\bibnamefont
  {Jurcevic}}, \bibinfo {author} {\bibfnamefont {B.~P.}\ \bibnamefont
  {Lanyon}}, \bibinfo {author} {\bibfnamefont {P.}~\bibnamefont {Hauke}},
  \bibinfo {author} {\bibfnamefont {C.}~\bibnamefont {Hempel}}, \bibinfo
  {author} {\bibfnamefont {P.}~\bibnamefont {Zoller}}, \bibinfo {author}
  {\bibfnamefont {R.}~\bibnamefont {Blatt}}, \ and\ \bibinfo {author}
  {\bibfnamefont {C.~F.}\ \bibnamefont {Roos}},\ }\href {\doibase
  10.1038/nature13461} {\bibfield  {journal} {\bibinfo  {journal} {Nature}\
  }\textbf {\bibinfo {volume} {511}},\ \bibinfo {pages} {202} (\bibinfo {year}
  {2014})}\BibitemShut {NoStop}%
\bibitem [{\citenamefont {Britton}\ \emph {et~al.}(2012)\citenamefont
  {Britton}, \citenamefont {Sawyer}, \citenamefont {Keith}, \citenamefont
  {Wang}, \citenamefont {Freericks}, \citenamefont {Uys}, \citenamefont
  {Biercuk},\ and\ \citenamefont {Bollinger}}]{britton12}%
  \BibitemOpen
  \bibfield  {author} {\bibinfo {author} {\bibfnamefont {J.~W.}\ \bibnamefont
  {Britton}}, \bibinfo {author} {\bibfnamefont {B.~C.}\ \bibnamefont {Sawyer}},
  \bibinfo {author} {\bibfnamefont {A.~C.}\ \bibnamefont {Keith}}, \bibinfo
  {author} {\bibfnamefont {C.-C.~J.}\ \bibnamefont {Wang}}, \bibinfo {author}
  {\bibfnamefont {J.~K.}\ \bibnamefont {Freericks}}, \bibinfo {author}
  {\bibfnamefont {H.}~\bibnamefont {Uys}}, \bibinfo {author} {\bibfnamefont
  {M.~J.}\ \bibnamefont {Biercuk}}, \ and\ \bibinfo {author} {\bibfnamefont
  {J.~J.}\ \bibnamefont {Bollinger}},\ }\href@noop {} {\bibfield  {journal}
  {\bibinfo  {journal} {Nature}\ }\textbf {\bibinfo {volume} {484}},\ \bibinfo
  {pages} {489} (\bibinfo {year} {2012})}\BibitemShut {NoStop}%
\bibitem [{\citenamefont {Bohnet}\ \emph {et~al.}(2016)\citenamefont {Bohnet},
  \citenamefont {Sawyer}, \citenamefont {Britton}, \citenamefont {Wall},
  \citenamefont {Rey}, \citenamefont {Foss-Feig},\ and\ \citenamefont
  {Bollinger}}]{bohnet15}%
  \BibitemOpen
  \bibfield  {author} {\bibinfo {author} {\bibfnamefont {J.~G.}\ \bibnamefont
  {Bohnet}}, \bibinfo {author} {\bibfnamefont {B.~C.}\ \bibnamefont {Sawyer}},
  \bibinfo {author} {\bibfnamefont {J.~W.}\ \bibnamefont {Britton}}, \bibinfo
  {author} {\bibfnamefont {M.~L.}\ \bibnamefont {Wall}}, \bibinfo {author}
  {\bibfnamefont {A.~M.}\ \bibnamefont {Rey}}, \bibinfo {author} {\bibfnamefont
  {M.}~\bibnamefont {Foss-Feig}}, \ and\ \bibinfo {author} {\bibfnamefont
  {J.~J.}\ \bibnamefont {Bollinger}},\ }\href {\doibase
  10.1126/science.aad9958} {\bibfield  {journal} {\bibinfo  {journal}
  {Science}\ }\textbf {\bibinfo {volume} {352}},\ \bibinfo {pages} {1297}
  (\bibinfo {year} {2016})}\BibitemShut {NoStop}%
\bibitem [{\citenamefont {{Zhang}}\ \emph {et~al.}(2017)\citenamefont
  {{Zhang}}, \citenamefont {{Pagano}}, \citenamefont {{Hess}}, \citenamefont
  {{Kyprianidis}}, \citenamefont {{Becker}}, \citenamefont {{Kaplan}},
  \citenamefont {{Gorshkov}}, \citenamefont {{Gong}},\ and\ \citenamefont
  {{Monroe}}}]{Zhang2017}%
  \BibitemOpen
  \bibfield  {author} {\bibinfo {author} {\bibfnamefont {J.}~\bibnamefont
  {{Zhang}}}, \bibinfo {author} {\bibfnamefont {G.}~\bibnamefont {{Pagano}}},
  \bibinfo {author} {\bibfnamefont {P.~W.}\ \bibnamefont {{Hess}}}, \bibinfo
  {author} {\bibfnamefont {A.}~\bibnamefont {{Kyprianidis}}}, \bibinfo {author}
  {\bibfnamefont {P.}~\bibnamefont {{Becker}}}, \bibinfo {author}
  {\bibfnamefont {H.}~\bibnamefont {{Kaplan}}}, \bibinfo {author}
  {\bibfnamefont {A.~V.}\ \bibnamefont {{Gorshkov}}}, \bibinfo {author}
  {\bibfnamefont {Z.-X.}\ \bibnamefont {{Gong}}}, \ and\ \bibinfo {author}
  {\bibfnamefont {C.}~\bibnamefont {{Monroe}}},\ }\href@noop {} {\bibfield
  {journal} {\bibinfo  {journal} {ArXiv e-prints}\ } (\bibinfo {year}
  {2017})},\ \Eprint {http://arxiv.org/abs/1708.01044} {arXiv:1708.01044
  [quant-ph]} \BibitemShut {NoStop}%
\bibitem [{\citenamefont {N{\"a}gerl}\ \emph {et~al.}(1999)\citenamefont
  {N{\"a}gerl}, \citenamefont {Leibfried}, \citenamefont {Rohde}, \citenamefont
  {Thalhammer}, \citenamefont {Eschner}, \citenamefont {Schmidt-Kaler},\ and\
  \citenamefont {Blatt}}]{naegerl99}%
  \BibitemOpen
  \bibfield  {author} {\bibinfo {author} {\bibfnamefont {H.~C.}\ \bibnamefont
  {N{\"a}gerl}}, \bibinfo {author} {\bibfnamefont {D.}~\bibnamefont
  {Leibfried}}, \bibinfo {author} {\bibfnamefont {H.}~\bibnamefont {Rohde}},
  \bibinfo {author} {\bibfnamefont {G.}~\bibnamefont {Thalhammer}}, \bibinfo
  {author} {\bibfnamefont {J.}~\bibnamefont {Eschner}}, \bibinfo {author}
  {\bibfnamefont {F.}~\bibnamefont {Schmidt-Kaler}}, \ and\ \bibinfo {author}
  {\bibfnamefont {R.}~\bibnamefont {Blatt}},\ }\href {\doibase
  10.1103/PhysRevA.60.145} {\bibfield  {journal} {\bibinfo  {journal} {Phys.
  Rev. A}\ }\textbf {\bibinfo {volume} {60}},\ \bibinfo {pages} {145} (\bibinfo
  {year} {1999})}\BibitemShut {NoStop}%
\bibitem [{\citenamefont {Smith}\ \emph {et~al.}(2016)\citenamefont {Smith},
  \citenamefont {Lee}, \citenamefont {Richerme}, \citenamefont {Neyenhuis},
  \citenamefont {Hess}, \citenamefont {Hauke}, \citenamefont {Heyl},
  \citenamefont {Huse},\ and\ \citenamefont {Monroe}}]{Smith2016}%
  \BibitemOpen
  \bibfield  {author} {\bibinfo {author} {\bibfnamefont {J.}~\bibnamefont
  {Smith}}, \bibinfo {author} {\bibfnamefont {A.}~\bibnamefont {Lee}}, \bibinfo
  {author} {\bibfnamefont {P.}~\bibnamefont {Richerme}}, \bibinfo {author}
  {\bibfnamefont {B.}~\bibnamefont {Neyenhuis}}, \bibinfo {author}
  {\bibfnamefont {P.~W.}\ \bibnamefont {Hess}}, \bibinfo {author}
  {\bibfnamefont {P.}~\bibnamefont {Hauke}}, \bibinfo {author} {\bibfnamefont
  {M.}~\bibnamefont {Heyl}}, \bibinfo {author} {\bibfnamefont {D.~A.}\
  \bibnamefont {Huse}}, \ and\ \bibinfo {author} {\bibfnamefont
  {C.}~\bibnamefont {Monroe}},\ }\href {http://dx.doi.org/10.1038/nphys3783}
  {\bibfield  {journal} {\bibinfo  {journal} {Nat Phys}\ }\textbf {\bibinfo
  {volume} {12}},\ \bibinfo {pages} {907} (\bibinfo {year} {2016})}\BibitemShut
  {NoStop}%
\bibitem [{\citenamefont {Roos}\ \emph {et~al.}(2004)\citenamefont {Roos},
  \citenamefont {Riebe}, \citenamefont {H{\"a}ffner}, \citenamefont
  {H{\"a}nsel}, \citenamefont {Benhelm}, \citenamefont {Lancaster},
  \citenamefont {Becher}, \citenamefont {Schmidt-Kaler},\ and\ \citenamefont
  {Blatt}}]{roos04}%
  \BibitemOpen
  \bibfield  {author} {\bibinfo {author} {\bibfnamefont {C.~F.}\ \bibnamefont
  {Roos}}, \bibinfo {author} {\bibfnamefont {M.}~\bibnamefont {Riebe}},
  \bibinfo {author} {\bibfnamefont {H.}~\bibnamefont {H{\"a}ffner}}, \bibinfo
  {author} {\bibfnamefont {W.}~\bibnamefont {H{\"a}nsel}}, \bibinfo {author}
  {\bibfnamefont {J.}~\bibnamefont {Benhelm}}, \bibinfo {author} {\bibfnamefont
  {G.~P.~T.}\ \bibnamefont {Lancaster}}, \bibinfo {author} {\bibfnamefont
  {C.}~\bibnamefont {Becher}}, \bibinfo {author} {\bibfnamefont
  {F.}~\bibnamefont {Schmidt-Kaler}}, \ and\ \bibinfo {author} {\bibfnamefont
  {R.}~\bibnamefont {Blatt}},\ }\href {\doibase 10.1126/science.1097522}
  {\bibfield  {journal} {\bibinfo  {journal} {Science}\ }\textbf {\bibinfo
  {volume} {304}},\ \bibinfo {pages} {1478} (\bibinfo {year}
  {2004})}\BibitemShut {NoStop}%
\bibitem [{\citenamefont {{Jurcevic}}\ \emph {et~al.}(2016)\citenamefont
  {{Jurcevic}}, \citenamefont {{Shen}}, \citenamefont {{Hauke}}, \citenamefont
  {{Maier}}, \citenamefont {{Brydges}}, \citenamefont {{Hempel}}, \citenamefont
  {{Lanyon}}, \citenamefont {{Heyl}}, \citenamefont {{Blatt}},\ and\
  \citenamefont {{Roos}}}]{Jurcevic2017}%
  \BibitemOpen
  \bibfield  {author} {\bibinfo {author} {\bibfnamefont {P.}~\bibnamefont
  {{Jurcevic}}}, \bibinfo {author} {\bibfnamefont {H.}~\bibnamefont {{Shen}}},
  \bibinfo {author} {\bibfnamefont {P.}~\bibnamefont {{Hauke}}}, \bibinfo
  {author} {\bibfnamefont {C.}~\bibnamefont {{Maier}}}, \bibinfo {author}
  {\bibfnamefont {T.}~\bibnamefont {{Brydges}}}, \bibinfo {author}
  {\bibfnamefont {C.}~\bibnamefont {{Hempel}}}, \bibinfo {author}
  {\bibfnamefont {B.~P.}\ \bibnamefont {{Lanyon}}}, \bibinfo {author}
  {\bibfnamefont {M.}~\bibnamefont {{Heyl}}}, \bibinfo {author} {\bibfnamefont
  {R.}~\bibnamefont {{Blatt}}}, \ and\ \bibinfo {author} {\bibfnamefont
  {C.~F.}\ \bibnamefont {{Roos}}},\ }\href@noop {} {\bibfield  {journal}
  {\bibinfo  {journal} {ArXiv e-prints}\ } (\bibinfo {year} {2016})},\ \Eprint
  {http://arxiv.org/abs/1612.06902} {arXiv:1612.06902 [quant-ph]} \BibitemShut
  {NoStop}%
\bibitem [{\citenamefont {Gra{\ss}}\ \emph {et~al.}(2015)\citenamefont
  {Gra{\ss}}, \citenamefont {Muschik}, \citenamefont {Celi}, \citenamefont
  {Chhajlany},\ and\ \citenamefont {Lewenstein}}]{grass15}%
  \BibitemOpen
  \bibfield  {author} {\bibinfo {author} {\bibfnamefont {T.}~\bibnamefont
  {Gra{\ss}}}, \bibinfo {author} {\bibfnamefont {C.}~\bibnamefont {Muschik}},
  \bibinfo {author} {\bibfnamefont {A.}~\bibnamefont {Celi}}, \bibinfo {author}
  {\bibfnamefont {R.~W.}\ \bibnamefont {Chhajlany}}, \ and\ \bibinfo {author}
  {\bibfnamefont {M.}~\bibnamefont {Lewenstein}},\ }\href {\doibase
  10.1103/PhysRevA.91.063612} {\bibfield  {journal} {\bibinfo  {journal} {Phys.
  Rev. A}\ }\textbf {\bibinfo {volume} {91}},\ \bibinfo {pages} {063612}
  (\bibinfo {year} {2015})}\BibitemShut {NoStop}%
\bibitem [{\citenamefont {Dunlap}\ and\ \citenamefont
  {Kenkre}(1986)}]{dunlap86}%
  \BibitemOpen
  \bibfield  {author} {\bibinfo {author} {\bibfnamefont {D.~H.}\ \bibnamefont
  {Dunlap}}\ and\ \bibinfo {author} {\bibfnamefont {V.~M.}\ \bibnamefont
  {Kenkre}},\ }\href {\doibase 10.1103/PhysRevB.34.3625} {\bibfield  {journal}
  {\bibinfo  {journal} {Phys. Rev. B}\ }\textbf {\bibinfo {volume} {34}},\
  \bibinfo {pages} {3625} (\bibinfo {year} {1986})}\BibitemShut {NoStop}%
\bibitem [{\citenamefont {Grossmann}\ \emph {et~al.}(1991)\citenamefont
  {Grossmann}, \citenamefont {Dittrich}, \citenamefont {Jung},\ and\
  \citenamefont {H{\"a}nggi}}]{haenggi91}%
  \BibitemOpen
  \bibfield  {author} {\bibinfo {author} {\bibfnamefont {F.}~\bibnamefont
  {Grossmann}}, \bibinfo {author} {\bibfnamefont {T.}~\bibnamefont {Dittrich}},
  \bibinfo {author} {\bibfnamefont {P.}~\bibnamefont {Jung}}, \ and\ \bibinfo
  {author} {\bibfnamefont {P.}~\bibnamefont {H{\"a}nggi}},\ }\href {\doibase
  10.1103/PhysRevLett.67.516} {\bibfield  {journal} {\bibinfo  {journal} {Phys.
  Rev. Lett.}\ }\textbf {\bibinfo {volume} {67}},\ \bibinfo {pages} {516}
  (\bibinfo {year} {1991})}\BibitemShut {NoStop}%
\bibitem [{\citenamefont {Holthaus}(1992)}]{holthaus92}%
  \BibitemOpen
  \bibfield  {author} {\bibinfo {author} {\bibfnamefont {M.}~\bibnamefont
  {Holthaus}},\ }\href {\doibase 10.1103/PhysRevLett.69.351} {\bibfield
  {journal} {\bibinfo  {journal} {Phys. Rev. Lett.}\ }\textbf {\bibinfo
  {volume} {69}},\ \bibinfo {pages} {351} (\bibinfo {year} {1992})}\BibitemShut
  {NoStop}%
\bibitem [{\citenamefont {Holthaus}\ and\ \citenamefont
  {Hone}(1993)}]{holthaus93}%
  \BibitemOpen
  \bibfield  {author} {\bibinfo {author} {\bibfnamefont {M.}~\bibnamefont
  {Holthaus}}\ and\ \bibinfo {author} {\bibfnamefont {D.}~\bibnamefont
  {Hone}},\ }\href {\doibase 10.1103/PhysRevB.47.6499} {\bibfield  {journal}
  {\bibinfo  {journal} {Phys. Rev. B}\ }\textbf {\bibinfo {volume} {47}},\
  \bibinfo {pages} {6499} (\bibinfo {year} {1993})}\BibitemShut {NoStop}%
\bibitem [{\citenamefont {Eckardt}\ \emph {et~al.}(2005)\citenamefont
  {Eckardt}, \citenamefont {Weiss},\ and\ \citenamefont {Holthaus}}]{andre05}%
  \BibitemOpen
  \bibfield  {author} {\bibinfo {author} {\bibfnamefont {A.}~\bibnamefont
  {Eckardt}}, \bibinfo {author} {\bibfnamefont {C.}~\bibnamefont {Weiss}}, \
  and\ \bibinfo {author} {\bibfnamefont {M.}~\bibnamefont {Holthaus}},\ }\href
  {\doibase 10.1103/PhysRevLett.95.260404} {\bibfield  {journal} {\bibinfo
  {journal} {Phys. Rev. Lett.}\ }\textbf {\bibinfo {volume} {95}},\ \bibinfo
  {pages} {260404} (\bibinfo {year} {2005})}\BibitemShut {NoStop}%
\bibitem [{\citenamefont {Eckardt}\ \emph {et~al.}(2010)\citenamefont
  {Eckardt}, \citenamefont {Hauke}, \citenamefont {Soltan-Panahi},
  \citenamefont {Becker}, \citenamefont {Sengstock},\ and\ \citenamefont
  {Lewenstein}}]{andre10}%
  \BibitemOpen
  \bibfield  {author} {\bibinfo {author} {\bibfnamefont {A.}~\bibnamefont
  {Eckardt}}, \bibinfo {author} {\bibfnamefont {P.}~\bibnamefont {Hauke}},
  \bibinfo {author} {\bibfnamefont {P.}~\bibnamefont {Soltan-Panahi}}, \bibinfo
  {author} {\bibfnamefont {C.}~\bibnamefont {Becker}}, \bibinfo {author}
  {\bibfnamefont {K.}~\bibnamefont {Sengstock}}, \ and\ \bibinfo {author}
  {\bibfnamefont {M.}~\bibnamefont {Lewenstein}},\ }\href
  {http://stacks.iop.org/0295-5075/89/i=1/a=10010} {\bibfield  {journal}
  {\bibinfo  {journal} {EPL (Europhysics Letters)}\ }\textbf {\bibinfo {volume}
  {89}},\ \bibinfo {pages} {10010} (\bibinfo {year} {2010})}\BibitemShut
  {NoStop}%
\bibitem [{\citenamefont {Lignier}\ \emph {et~al.}(2007)\citenamefont
  {Lignier}, \citenamefont {Sias}, \citenamefont {Ciampini}, \citenamefont
  {Singh}, \citenamefont {Zenesini}, \citenamefont {Morsch},\ and\
  \citenamefont {Arimondo}}]{lignier07}%
  \BibitemOpen
  \bibfield  {author} {\bibinfo {author} {\bibfnamefont {H.}~\bibnamefont
  {Lignier}}, \bibinfo {author} {\bibfnamefont {C.}~\bibnamefont {Sias}},
  \bibinfo {author} {\bibfnamefont {D.}~\bibnamefont {Ciampini}}, \bibinfo
  {author} {\bibfnamefont {Y.}~\bibnamefont {Singh}}, \bibinfo {author}
  {\bibfnamefont {A.}~\bibnamefont {Zenesini}}, \bibinfo {author}
  {\bibfnamefont {O.}~\bibnamefont {Morsch}}, \ and\ \bibinfo {author}
  {\bibfnamefont {E.}~\bibnamefont {Arimondo}},\ }\href {\doibase
  10.1103/PhysRevLett.99.220403} {\bibfield  {journal} {\bibinfo  {journal}
  {Phys. Rev. Lett.}\ }\textbf {\bibinfo {volume} {99}},\ \bibinfo {pages}
  {220403} (\bibinfo {year} {2007})}\BibitemShut {NoStop}%
\bibitem [{\citenamefont {Struck}\ \emph {et~al.}(2011)\citenamefont {Struck},
  \citenamefont {{\"O}lschl{\"a}ger}, \citenamefont {{Le Targat}},
  \citenamefont {Soltan-Panahi}, \citenamefont {Eckardt}, \citenamefont
  {Lewenstein}, \citenamefont {Windpassinger},\ and\ \citenamefont
  {Sengstock}}]{struck11}%
  \BibitemOpen
  \bibfield  {author} {\bibinfo {author} {\bibfnamefont {J.}~\bibnamefont
  {Struck}}, \bibinfo {author} {\bibfnamefont {C.}~\bibnamefont
  {{\"O}lschl{\"a}ger}}, \bibinfo {author} {\bibfnamefont {R.}~\bibnamefont
  {{Le Targat}}}, \bibinfo {author} {\bibfnamefont {P.}~\bibnamefont
  {Soltan-Panahi}}, \bibinfo {author} {\bibfnamefont {A.}~\bibnamefont
  {Eckardt}}, \bibinfo {author} {\bibfnamefont {M.}~\bibnamefont {Lewenstein}},
  \bibinfo {author} {\bibfnamefont {P.}~\bibnamefont {Windpassinger}}, \ and\
  \bibinfo {author} {\bibfnamefont {K.}~\bibnamefont {Sengstock}},\ }\href
  {\doibase 10.1126/science.1207239} {\bibfield  {journal} {\bibinfo  {journal}
  {Science}\ }\textbf {\bibinfo {volume} {333}},\ \bibinfo {pages} {996}
  (\bibinfo {year} {2011})}\BibitemShut {NoStop}%
\bibitem [{\citenamefont {Kitagawa}\ \emph {et~al.}(2010)\citenamefont
  {Kitagawa}, \citenamefont {Berg}, \citenamefont {Rudner},\ and\ \citenamefont
  {Demler}}]{kitagawa10}%
  \BibitemOpen
  \bibfield  {author} {\bibinfo {author} {\bibfnamefont {T.}~\bibnamefont
  {Kitagawa}}, \bibinfo {author} {\bibfnamefont {E.}~\bibnamefont {Berg}},
  \bibinfo {author} {\bibfnamefont {M.}~\bibnamefont {Rudner}}, \ and\ \bibinfo
  {author} {\bibfnamefont {E.}~\bibnamefont {Demler}},\ }\href {\doibase
  10.1103/PhysRevB.82.235114} {\bibfield  {journal} {\bibinfo  {journal} {Phys.
  Rev. B}\ }\textbf {\bibinfo {volume} {82}},\ \bibinfo {pages} {235114}
  (\bibinfo {year} {2010})}\BibitemShut {NoStop}%
\bibitem [{\citenamefont {Jiang}\ \emph {et~al.}(2011)\citenamefont {Jiang},
  \citenamefont {Kitagawa}, \citenamefont {Alicea}, \citenamefont {Akhmerov},
  \citenamefont {Pekker}, \citenamefont {Refael}, \citenamefont {Cirac},
  \citenamefont {Demler}, \citenamefont {Lukin},\ and\ \citenamefont
  {Zoller}}]{jiang11}%
  \BibitemOpen
  \bibfield  {author} {\bibinfo {author} {\bibfnamefont {L.}~\bibnamefont
  {Jiang}}, \bibinfo {author} {\bibfnamefont {T.}~\bibnamefont {Kitagawa}},
  \bibinfo {author} {\bibfnamefont {J.}~\bibnamefont {Alicea}}, \bibinfo
  {author} {\bibfnamefont {A.~R.}\ \bibnamefont {Akhmerov}}, \bibinfo {author}
  {\bibfnamefont {D.}~\bibnamefont {Pekker}}, \bibinfo {author} {\bibfnamefont
  {G.}~\bibnamefont {Refael}}, \bibinfo {author} {\bibfnamefont {J.~I.}\
  \bibnamefont {Cirac}}, \bibinfo {author} {\bibfnamefont {E.}~\bibnamefont
  {Demler}}, \bibinfo {author} {\bibfnamefont {M.~D.}\ \bibnamefont {Lukin}}, \
  and\ \bibinfo {author} {\bibfnamefont {P.}~\bibnamefont {Zoller}},\ }\href
  {\doibase 10.1103/PhysRevLett.106.220402} {\bibfield  {journal} {\bibinfo
  {journal} {Phys. Rev. Lett.}\ }\textbf {\bibinfo {volume} {106}},\ \bibinfo
  {pages} {220402} (\bibinfo {year} {2011})}\BibitemShut {NoStop}%
\bibitem [{\citenamefont {Lindner}\ \emph {et~al.}(2011)\citenamefont
  {Lindner}, \citenamefont {Refael},\ and\ \citenamefont
  {Galitski}}]{lindner11}%
  \BibitemOpen
  \bibfield  {author} {\bibinfo {author} {\bibfnamefont {N.~H.}\ \bibnamefont
  {Lindner}}, \bibinfo {author} {\bibfnamefont {G.}~\bibnamefont {Refael}}, \
  and\ \bibinfo {author} {\bibfnamefont {V.}~\bibnamefont {Galitski}},\ }\href
  {\doibase 10.1038/nphys1926} {\bibfield  {journal} {\bibinfo  {journal} {Nat.
  Phys.}\ }\textbf {\bibinfo {volume} {7}},\ \bibinfo {pages} {490} (\bibinfo
  {year} {2011})},\ \bibinfo {note} {10.1038/nphys1926}\BibitemShut {NoStop}%
\bibitem [{\citenamefont {Bermudez}\ \emph {et~al.}(2011)\citenamefont
  {Bermudez}, \citenamefont {Schaetz},\ and\ \citenamefont
  {Porras}}]{bermudez11}%
  \BibitemOpen
  \bibfield  {author} {\bibinfo {author} {\bibfnamefont {A.}~\bibnamefont
  {Bermudez}}, \bibinfo {author} {\bibfnamefont {T.}~\bibnamefont {Schaetz}}, \
  and\ \bibinfo {author} {\bibfnamefont {D.}~\bibnamefont {Porras}},\ }\href
  {\doibase 10.1103/PhysRevLett.107.150501} {\bibfield  {journal} {\bibinfo
  {journal} {Phys. Rev. Lett.}\ }\textbf {\bibinfo {volume} {107}},\ \bibinfo
  {pages} {150501} (\bibinfo {year} {2011})}\BibitemShut {NoStop}%
\bibitem [{\citenamefont {Gra\ss}\ \emph {et~al.}(2016)\citenamefont {Gra\ss},
  \citenamefont {Lewenstein},\ and\ \citenamefont {Bermudez}}]{grass152}%
  \BibitemOpen
  \bibfield  {author} {\bibinfo {author} {\bibfnamefont {T.}~\bibnamefont
  {Gra\ss}}, \bibinfo {author} {\bibfnamefont {M.}~\bibnamefont {Lewenstein}},
  \ and\ \bibinfo {author} {\bibfnamefont {A.}~\bibnamefont {Bermudez}},\
  }\href {http://stacks.iop.org/1367-2630/18/i=3/a=033011} {\bibfield
  {journal} {\bibinfo  {journal} {New Journal of Physics}\ }\textbf {\bibinfo
  {volume} {18}},\ \bibinfo {pages} {033011} (\bibinfo {year}
  {2016})}\BibitemShut {NoStop}%
\bibitem [{\citenamefont {Wang}\ and\ \citenamefont
  {Freericks}(2012)}]{wang12}%
  \BibitemOpen
  \bibfield  {author} {\bibinfo {author} {\bibfnamefont {C.-C.~J.}\
  \bibnamefont {Wang}}\ and\ \bibinfo {author} {\bibfnamefont {J.~K.}\
  \bibnamefont {Freericks}},\ }\href {\doibase 10.1103/PhysRevA.86.032329}
  {\bibfield  {journal} {\bibinfo  {journal} {Phys. Rev. A}\ }\textbf {\bibinfo
  {volume} {86}},\ \bibinfo {pages} {032329} (\bibinfo {year}
  {2012})}\BibitemShut {NoStop}%
\bibitem [{\citenamefont {Wall}\ \emph {et~al.}(2017)\citenamefont {Wall},
  \citenamefont {Safavi-Naini},\ and\ \citenamefont {Rey}}]{Wall2017}%
  \BibitemOpen
  \bibfield  {author} {\bibinfo {author} {\bibfnamefont {M.~L.}\ \bibnamefont
  {Wall}}, \bibinfo {author} {\bibfnamefont {A.}~\bibnamefont {Safavi-Naini}},
  \ and\ \bibinfo {author} {\bibfnamefont {A.~M.}\ \bibnamefont {Rey}},\ }\href
  {\doibase 10.1103/PhysRevA.95.013602} {\bibfield  {journal} {\bibinfo
  {journal} {Phys. Rev. A}\ }\textbf {\bibinfo {volume} {95}},\ \bibinfo
  {pages} {013602} (\bibinfo {year} {2017})}\BibitemShut {NoStop}%
\bibitem [{Note1()}]{Note1}%
  \BibitemOpen
  \bibinfo {note} {Here, for simplicity we have assumed $B_0$ sufficiently
  larger than the Ising couplings $J_{ij}$ such that the net effect of driving
  is equivalent to the driving of a XX model by the time-dependent magnetic
  field $H_{B1}$. More generally, the condition to achieve an effective XX
  model with the complex coupling as in Eq. \protect \textup {\hbox
  {\mathsurround \z@ \protect \normalfont (\ignorespaces \ref {jeff}\unskip
  \@@italiccorr )}} is $\DOTSI \intop \ilimits@ _0^{\protect \mathaccentV
  {tilde}07ET} \protect \qopname \relax o{exp}[2 i (2 B_0 t + \mu _i(t) +\mu
  _j(t))] =0$, where $\protect \mathaccentV {tilde}07ET$ is the period of $ B_0
  t + \mu _i(t)$.}\BibitemShut {Stop}%
\end{thebibliography}

%

\end{document}